\documentclass[aps,pra,english,twocolumn,letterpaper,notitlepage,superscriptaddress]{revtex4-2}

\usepackage[T1]{fontenc}
\usepackage{graphicx}
\usepackage{epstopdf}
\usepackage{amssymb}
\usepackage{amsmath}
\usepackage[urlcolor=blue,hyperindex,colorlinks,bookmarks=true,linkcolor=red,citecolor=black]{hyperref}
\usepackage[normalem]{ulem}
\usepackage{color}
\usepackage{rotating}
\usepackage{physics}
\usepackage{bm}
\usepackage{soul}
\usepackage{enumitem}

\usepackage[english]{babel}

\def\iden{\hat{\mathbb{I}}}

\newboolean{ShowComments}
\setboolean{ShowComments}{false}

\definecolor{cpcolor}{rgb}{1,0,0}
\definecolor{mscolor}{rgb}{0,0.5,0.5}
\definecolor{lgcolor}{rgb}{0,0.0,1.0}

\begin{document}

\title{Freedom of mixer rotation-axis improves performance in the quantum approximate optimization algorithm}

\author{L.~C.~G.~Govia}
\affiliation{Quantum Engineering and Computing, Raytheon BBN Technologies, 10 Moulton St., Cambridge, MA 02138, USA}
\email{lcggovia@gmail.com}
\author{C. Poole}
\affiliation{Department of Physics, University of Wisconsin-Madison, Madison, WI 53706, USA}
\author{M. Saffman}
\affiliation{Department of Physics, University of Wisconsin-Madison, Madison, WI 53706, USA}
\affiliation{ColdQuanta, Inc.,  111 N. Fairchild St., Madison, WI 53703, USA}
\author{H.~K.~Krovi}
\affiliation{Quantum Engineering and Computing, Raytheon BBN Technologies, 10 Moulton St., Cambridge, MA 02138, USA}

\begin{abstract}
  Variational quantum algorithms such as the quantum approximate optimization algorithm (QAOA) are particularly attractive candidates for implementation on near-term quantum processors. As hardware realities such as error and qubit connectivity will constrain achievable circuit depth in the near future, new ways to achieve high-performance at low depth are of great interest. In this work, we present a modification to QAOA that adds additional variational parameters in the form of freedom of the rotation-axis in the $XY$-plane of the mixer Hamiltonian. Via numerical simulation, we show that this leads to a drastic performance improvement over standard QAOA at finding solutions to the MAXCUT problem on graphs of up to 7 qubits. Furthermore, we explore the Z-phase error mitigation properties of our modified ansatz, its performance under a realistic error model for a neutral atom quantum processor, and the class of problems it can solve in a single round.
\end{abstract}

\maketitle

\section{Introduction}

Variational quantum algorithms (VQAs) \cite{Cerezo2020,Biamonte2021} show much promise for realizing potential quantum advantage in the noisy-intermediate-scale quantum (NISQ) era \cite{Preskill2018}. As opposed to standard quantum algorithms, where a quantum circuit is designed to solve a particular problem, VQAs leverage classical co-processing and optimization to find a circuit that effectively performs the desired computation. While early demonstrations on small-scale quantum processors have been encouraging \cite{Kandala2017,Colless2018,Hempel2018,Kokail2019,Kandala2019,Havlicek2019,Zhu2019,Arute2020}, it remains an active area of research to develop new VQAs that satisfy the often competing requirements of increasing expressibility (exploring large portions of Hilbert space), reducing classical co-processing, and hardware efficiency. Given experimental realities such as error and qubit-connectivity, the number of variational parameters is often kept small to satisfy the latter two requirements at the cost of the former.

As such, there exists tremendous opportunity to improve VQA performance with carefully designed variational circuit ans\"atze that introduce additional parameters in a targeted way. Recent examples include QOCA \cite{Choquette2021}, which leverages quantum control inspired variational circuit design; ADAPT-VQE \cite{Grimsley2019,Tang2021}, which varies circuit gates/structure as well as free parameters; {\tt Rotoselect} \cite{Ostaszewski2021} and {\tt Fraxis} \cite{Watanabe2021}, which give discrete or continuous freedom to the single-qubit rotations in the variational circuit; and ab-QAOA \cite{Yu2021}, which adds a Pauli-Z component to the mixer Hamiltonian. Despite these recent advances, it remains an open question where additional variational freedom is best added to a structured VQA ansatz, such as one which uses the encoded problem Hamiltonian directly.

In this work, we introduce a modified version of the popular quantum approximate optimization algorithm (QAOA) \cite{Farhi:2014wk}. Our modification is intended to both expand the region of Hilbert space explored, and mitigate Z-phase error, while remaining as hardware-efficient as standard QAOA. We demonstrate through numerical simulation on up to 7 qubits that our modified QAOA drastically outperforms standard QAOA, especially at low depth. This performance improvement occurs for all error models that we consider, including a physical error model for a neutral atom quantum processor with Rydberg state mediated gates. We believe our modified ansatz can find immediate implementation in current generation hardware platforms aiming to solve optimization problems with QAOA-like VQAs.

This article is organized as follows. In section \ref{sec:FAM} we review standard QAOA and introduce our modified version. In section \ref{sec:Zphase} we discuss how our ansatz mitigates Z-phase error, and in section \ref{sec:sims} we present the results of our numerical simulations. Finally, in section \ref{sec:exact} we elucidate the class of problem Hamiltonians that can be solved in a single round with our ansatz, and in \ref{sec:conc} we make our concluding remarks.

\section{Free Axis Mixer QAOA}
\label{sec:FAM}

One of the most well-studied VQAs is the quantum approximate optimization algorithm (QAOA) \cite{Farhi:2014wk}, which has a variational circuit that takes the form
\begin{align}
     \hat{U}_{\rm QAOA} = \prod_{k=1}^p \exp(-i\beta_k\sum_{n=1}^N\hat{X}_n)\exp(-i\gamma_k\hat{H}_Z),
\end{align}
where $\hat{H}_Z$ is the Ising-model encoding of the problem, and $\hat{X}_n$ is shorthand for Pauli-X on qubit $n$ and identity on all others. Inspired by QAOA, the quantum alternating operator ansatz \cite{Hadfield:2019ul} can be thought of as a generalization, and consists of a variational circuit of the form
\begin{align}
    \hat{U}_{\rm qaoa} = \prod_{k=1}^p \hat{U}_{M}(\boldsymbol{\beta}_k)\hat{U}_Z(\boldsymbol{\gamma}_k),
\end{align}
where $\hat{U}_M(\boldsymbol{\beta}_k)$ and $\hat{U}_Z(\boldsymbol{\gamma}_k)$ are referred to as the mixer and problem unitaries, and are functions of the variational parameters $\boldsymbol{\beta}_k$ and $\boldsymbol{\gamma}_k$. These unitaries can often be described in terms of generating Hamiltonians $\hat{H}_M$ and $\hat{H}_Z$ that are functions of the variational parameters, such that
\begin{align}
    \hat{U}_{M}(\boldsymbol{\beta}_k) = e^{-i\hat{H}_M(\boldsymbol{\beta}_k)},~\hat{U}_{Z}(\boldsymbol{\gamma}_k) = e^{-i\hat{H}_Z(\boldsymbol{\gamma}_k)}.
\end{align}
In fact, the mixer and problem Hamiltonian typically have the form $\hat{H}_M(\boldsymbol{\beta}_k) = \beta_k\hat{H}_M$, and $\hat{H}_Z(\boldsymbol{\gamma}_k) = \gamma_k\hat{H}_Z$, such that there is one variational parameter per-unitary, per-round. As in QAOA, $\hat{H}_Z$ is typically the Ising-model encoding of the problem to be solved, and is thus diagonal in the Pauli-basis. It may additionally encode constraints on the optimization \cite{Hadfield:2019ul}.

Developments beyond the Pauli-X mixer have with few exceptions maintained the paradigm of a single mixer free parameter $\beta_k$ per round. As reviewed in great detail in Ref.~\cite{Hadfield:2019ul}, examples of modified mixers that have been proposed are those that act on only a subset of qubits (such as due to graph considerations \cite{McClean2020}), or introduce entangling operations, often to encode hard constraints \cite{Fingerhuth2018,Wang2020}. To the best of our knowledge, the exceptions to the single $\beta_k$ paradigm are Refs.~\cite{Farhi2017,Bapat2019}, which consider separate mixer angles for each qubit, and Ref.~\cite{Yu2021}, which adds a free parameter for each qubit to parameterize a Pauli-Z component in the mixer Hamiltonian.

We introduce a modified mixer whose design was foremost informed by practical hardware-efficiency. At round $k$, our mixer Hamiltonian has the form
\begin{align}
    \hat{H}_M^{(k)}\left(\theta_{n}^k\right) = \sum_{n=1}^N\left(\cos(\theta_{n}^k)\hat{X}_n - \sin(\theta_{n}^k)\hat{Y}_n\right), \label{eqn:FAMQAOA}
\end{align}
which incorporates additional variational parameters $\theta^{k}_n$. This mixer describes a rotation of qubit $n$ in round $k$ about an axis in the $XY$-plane defined by the angle $\theta^{k}_n$, and for that reason we call it the free axis mixer quantum alternating operator ansatz (FAM-QAOA).

By expanding the possible single-qubit rotations of the mixer Hamiltonian to include rotations about any axis in the $XY$-plane, FAM-QAOA expands the possible states that can be reached by the mixer operation. Modification of the rotation-axis of a single-qubit drive is a trivial operation in most leading platforms for quantum computing (e.g.~neutral atoms \cite{Xia2015}, superconducting qubits \cite{McKay2017}, or trapped ions \cite{Debnath2016}) which requires little to no additional experimental complexity or overhead.

We note that the free axis mixer of FAM-QAOA is related to the free axis selection of {\tt Fraxis} \cite{Watanabe2021}. The distinctions are that i) FAM-QAOA is specific to a variational circuit with QAOA structure, while {\tt Fraxis} considers a parameterized quantum circuit with fixed entangling gates and variational single qubit gates, and ii) FAM-QAOA restricts the axis of rotation to the $XY$-plane, and so has one less free parameter per qubit. It is also related to ab-QAOA \cite{Yu2021}, but FAM-QAOA adds a Pauli-Y component to the mixer that can be qubit- and round-dependent, while ab-QAOA adds a qubit-dependent Pauli-Z component. Other work has also considered QAOA with mixer rotations not aligned with the $X$-axis \cite{Egger2020}, but without the rotation-axis as a variational parameter.

In general, FAM-QAOA is a family of ans\"atze with a tunable number of additional variational parameters. As written, Eq.~\eqref{eqn:FAMQAOA} introduces $p\times N$ angles: one per-round, per-qubit. We dub this $pN$-FAM. We can restrict the number of free parameters by imposing that the angles be fixed per-round but variable per-qubit ($N$-FAM), fixed per-qubit but variable per-round ($p$-FAM), or a single angle defining a global axis of rotation ($1$-FAM).

For $N$-FAM and $1$-FAM, we also introduce the possibility of linearly scaling the angle defining the axis of rotation on a per-round basis. For $N$-FAM this would take the form
\begin{align}
    \hat{H}_{M,~{\rm scaled}}^{(k)} = \sum_{n=1}^N\left(\cos(k\theta_{n})\hat{X}_n - \sin(k\theta_{n})\hat{Y}_n\right). \label{eqn:FAMQAOAscaled}
\end{align}
In the next section we will motivate FAM-QAOA as an approach to mitigate the impact of coherent single-qubit Z-phase errors. For such an error model this linear scaling of the axis of rotation counteracts accumulation of a Z-phase error that is fixed per-round.

We will explore the performance of FAM-QAOA compared to standard QAOA, and how this performance depends on the number of free angles, i.e.~variational parameters, in the ansatz. Note that if the initial state (a product state of the same Pauli-X eigenstate for each qubit) can be modified as well, then $1$-FAM (without angle scaling) is identical to QAOA by a redefinition of the Pauli-X operator. However, we will not consider modification of the initial state in this work, such that $1$-FAM and QAOA are distinct.

\section{Cancellation of Z-Phase Error}
\label{sec:Zphase}

FAM-QAOA is in part motivated by the exact cancellation of static Z-phase errors made possible by the free axes of rotation. To understand how this is possible, we first rewrite the FAM-QAOA circuit as
\begin{align}
     \hat{U}_{\rm FAM} = \prod_{k=1}^p e^{i\boldsymbol{\theta}^k.\mathbf{Z}}e^{-i\beta_k\hat{H}_X}e^{-i\boldsymbol{\theta}^k.\mathbf{Z}}e^{-i\gamma_k\hat{H}_Z},
\end{align}
where $\hat{H}_X = \sum_{n=1}^N\hat{X}_k$ is the usual Pauli-X mixer of QAOA, and
\begin{align}
    \boldsymbol{\theta}^k.\mathbf{Z} = \sum_{n=1}^{N}\theta_n^k\hat{Z}_n,
\end{align}
with $\hat{Z}_n$ shorthand for the $N$-qubit operator with Pauli-Z on qubit $n$ and identity on all others. Since Z-phase errors commute with the problem Hamiltonian, without loss of generality we can assume that any phase error happens after the problem unitary, described by the circuit
\begin{align}
     \hat{U}_{\rm FAM} = \prod_{k=1}^p e^{i\boldsymbol{\theta}^k.\mathbf{Z}}e^{-i\beta_k\hat{H}_X}e^{-i\boldsymbol{\theta}^k.\mathbf{Z}}e^{-i\boldsymbol{\phi}^k.\mathbf{Z}}e^{-i\gamma_k\hat{H}_Z},
\end{align}
where $\boldsymbol{\phi}^k$ is the vector of Z-phase errors on each qubit at round $k$. Inserting identity operators correctly, we can rewrite this circuit as
\begin{align}
     &\hat{U}_{\rm FAM} = \\ \nonumber& e^{-i \boldsymbol{\varphi}^p.\mathbf{Z}} \prod_{k=1}^p e^{i(\boldsymbol{\theta}^k+ \boldsymbol{\varphi}^k).\mathbf{Z}}e^{-i\beta_k\hat{H}_X}e^{-i(\boldsymbol{\theta}^k + \boldsymbol{\varphi}^k).\mathbf{Z}}e^{-i\gamma_k\hat{H}_Z}, \label{eqn:FAMerrorcircuit}
\end{align}
where $\boldsymbol{\varphi}^k = \sum_{j=1}^k \boldsymbol{\phi}^j$ is the element-wise sum of the Z-phase error vectors up to round $k$. In the circuit of Eq.~\eqref{eqn:FAMerrorcircuit}, it is straightforward to see that the Z-phase error can be perfectly cancelled by setting $\boldsymbol{\theta}^k = -\boldsymbol{\varphi}^k$ for each round. The final Z-phase gate with angles $\boldsymbol{\varphi}^p$ has no effect as we assume measurements are performed in the Pauli-Z basis.

If the Z-phase error is qubit- or round-independent, then full $pN$-FAM is not required for cancellation, but $p$-FAM or $N$-FAM would suffice. For round-independent error, the scaled version of $N$-FAM should be used to account for the accumulation of Z-phase error with circuit depth, such that $\boldsymbol{\theta}^k = k\boldsymbol{\theta}$ for the variational angles $\boldsymbol{\theta}$, as discussed previously.

Such exact cancellation is possible if the error is constant for each implementation of the FAM-QAOA circuit, i.e.~every time the full circuit is run (with potentially new $\beta_k$ and $\gamma_k$), the Z-phase error vectors $\boldsymbol{\phi}^k$ are the same. This is likely an unrealistic assumption for current hardware. However, as we will soon see, FAM-QAOA performs just as well when the Z-phase error is a function of the ansatz parameters $\gamma_k$, as would be the case if the error source was in the implementation of $\hat{H}_Z$.

\section{Numerical Simulation Results}
\label{sec:sims}

Our numerical simulations consider finding solutions to the MAXCUT problem \cite{Garey1979} on small graph sizes, which has the problem Hamiltonian
\begin{align}
    \hat{H}_Z = \sum_{(n,m)\in E} \frac{1}{2}\left(\hat{Z}_{n}\hat{Z}_{m} - \iden \right),
\end{align}
where $E$ is the set of edges defining the graph. The goal is to prepare a state $\ket{\psi}$ with the lowest expectation value (or cost), $C_\psi = \bra{\psi}\hat{H}_Z\ket{\psi}$, possible. We measure performance using the approximation ratio (AR), defined as the ratio of the expectation value of $\ket{\psi}$ and the ground-state expectation value, $\alpha = C_\psi/C_{\rm min}$, and the success probability (SP), which is the overlap between $\ket{\psi}$ and the (possibly degenerate) ground-state subspace of $\hat{H}_Z$.

The first set of simulations we present directly implement by matrix exponentiation the unitary matrices of the FAM-QAOA
\begin{align}
     \hat{U}_{\rm FAM} = \prod_{k=1}^p \exp(-i\beta_k\hat{H}_M^k\left(\theta_n^k\right))\exp(-i\gamma_k\hat{H}_Z), \label{eqn:fullFAM}
\end{align}
under several different Z-phase error models. We call these simulations ``hardware agnostic''. The second set of simulations use a circuit compiler to compile the FAM-QAOA unitary onto the structure and native gate-set of a neutral atom quantum processor. For these simulations we consider an error model for the neutral atom processor.

In the following subsections we present the results of our simulations. Further specifics of the simulations (e.g.~optimization algorithm, number of cold-starts, circuit compiler specifics) can be found in Appendix \ref{app:simdetails}.

\subsection{Hardware Agnostic MAXCUT}

\begin{figure}[t!]
    \centering
    \includegraphics[width=\columnwidth]{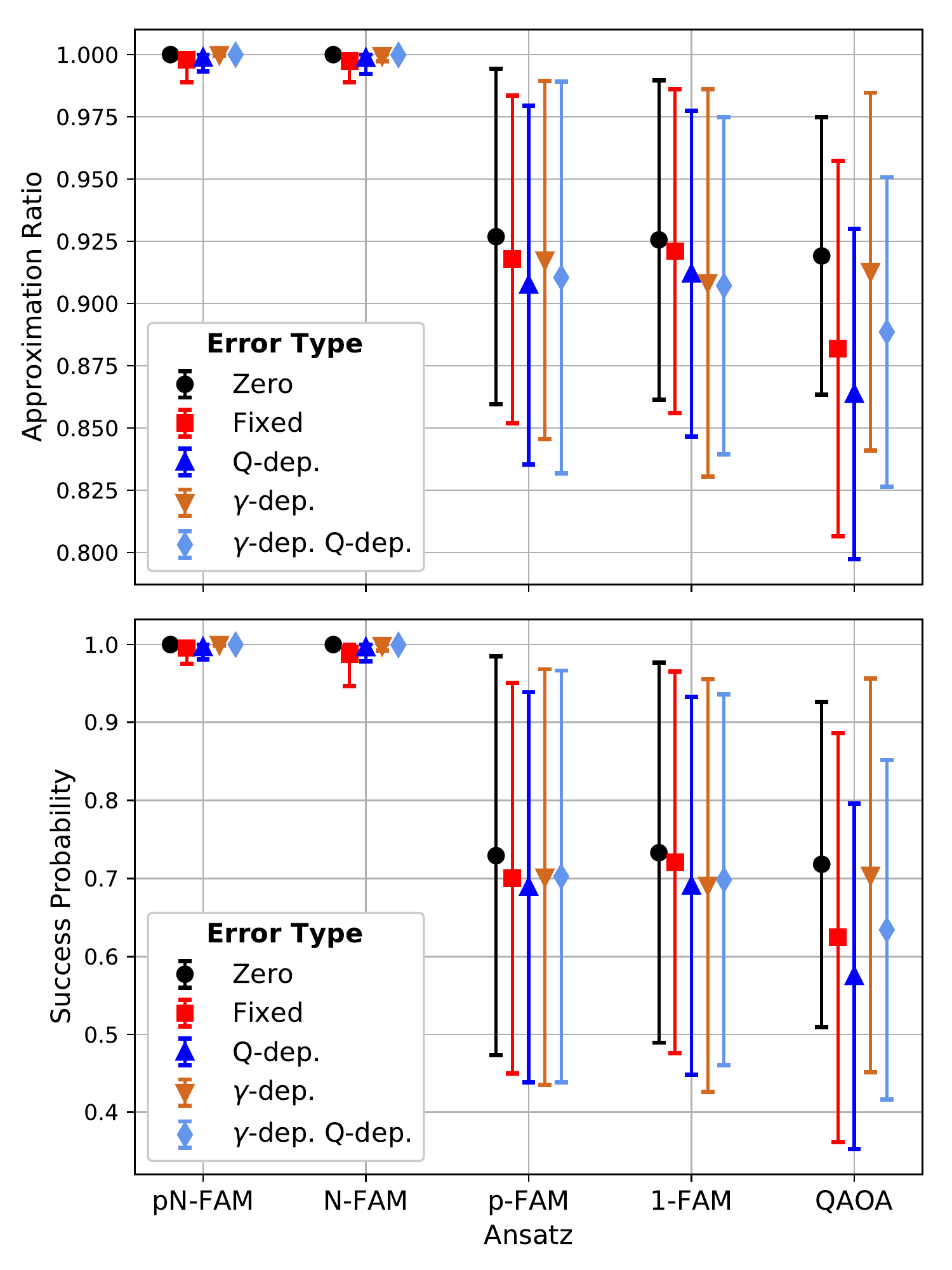}
    \caption{Average and error bars showing one standard deviation of the best-case (upper panel) approximation ratio and (lower panel) success probability at MAXCUT as a function of the number of free parameters in the FAM-QAOA for the five error types discussed in the main text. All 5-qubit connected graphs are considered and $p=2$, with $3$ cold-starts.}
    \label{fig:fig1}
\end{figure}

For the hardware agnostic simulations we consider error models that insert a Z-phase error gate
\begin{align}
    \hat{Z}\left(\boldsymbol{\phi}^k\right) = \exp(-i\sum_{n=1}^N \phi^k_n\hat{Z}_n),
\end{align}
after each application of the problem unitary. We consider five error models distinguished by the nature of the phase-error vector $\boldsymbol{\phi}^k$.
\begin{enumerate}[label=\roman*),leftmargin=*]
    \item Zero error: $\phi^k_n = 0~\forall~n~{\rm and}~k$.
    \item Fixed error: $\phi^k_n = \phi~\forall~n~{\rm and}~k$. Constant error on each qubit for each round.
    \item Qubit-dependent: $\phi^k_n = \phi_n~\forall~k$. Different error on each qubit, constant per round.
    \item $\gamma$-dependent: $\phi^k_n = \gamma_k\phi~\forall~n$. Constant $\gamma$-dependent error on each qubit.
    \item $\gamma$- and qubit-dependent: $\phi^k_n = \gamma_k\phi_n$. Different $\gamma$-dependent error on each qubit.
\end{enumerate}

For each of these error types, Fig.~\ref{fig:fig1} shows the best performance out of three repetitions with new initial conditions (``cold-starts'') of the FAM-QAOA versions, as well as standard QAOA for comparison. The symbols are the best-case AR and SP averaged over all connected $5$-qubit graphs and $p=2$, with the error bars showing one standard deviation of the performance distribution over the graphs. To ensure a fair comparison, for each ansatz we fix the number of cost function calls to the same constant, so that even the ansätze with more free parameters are not able to query the (simulated) quantum processor more often. See Appendix \ref{app:simdetails} for further details and Appendix \ref{app:moresim} for performance as a function of the number of function calls. As can be seen, $pN$- and $N$-FAM dramatically outperform all other ans\"atze, including standard QAOA, with near unit average AR and SP for all error models. The FAM-QAOA versions with less free parameters perform comparably to standard QAOA, and all FAM-QAOA versions show less performance-variation across error models.

\begin{figure}[t!]
    \centering
    \includegraphics[width=\columnwidth]{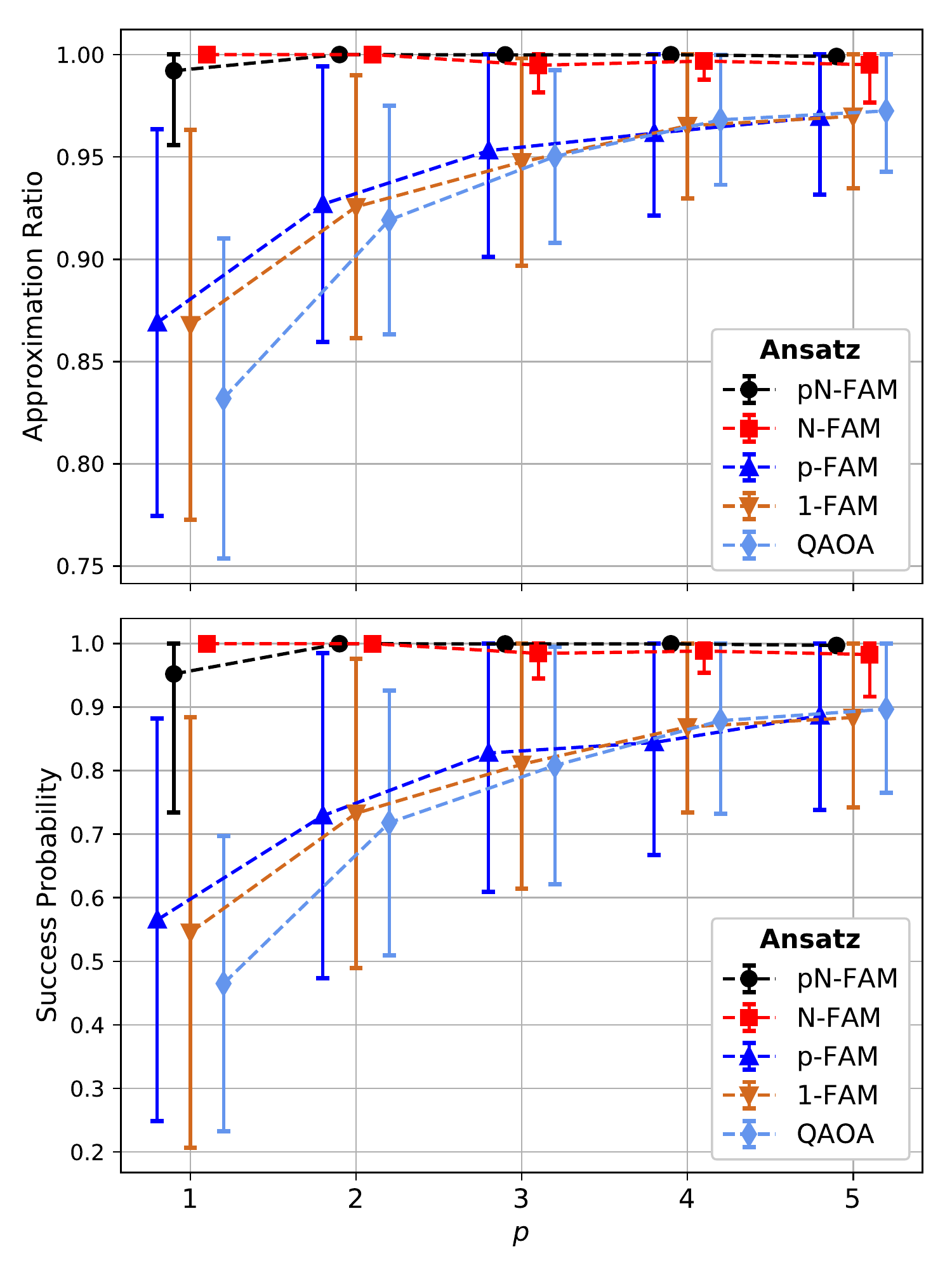}
    \caption{Average and error bars showing one standard deviation of the best-case (upper panel) approximation ratio and (lower panel) success probability at MAXCUT as a function of $p$ for all 5-qubit connected graphs with zero error. Note that the inconsistency between $pN$-FAM and $N$-FAM for $p=1$, where they are equivalent, is due to finite repetition statistics (see Appendix \ref{app:simdetails} for more details).}
    \label{fig:fig2}
\end{figure}

Remarkably, despite being designed to mitigate Z-phase error, FAM-QAOA outperforms standard QAOA even at zero error. To explore this further, we focus on zero error and examine performance as a function of $p$, again on all $5$-qubit connected graphs, as shown in Fig.~\ref{fig:fig2}. Unsurprisingly, $pN$- and $N$-FAM have near unit average AR and SP for $p\geq 2$, but intriguingly they only show a small degradation in performance for $p=1$. Also as expected, standard QAOA and the other versions of FAM-QAOA show steady improvement as $p$ increases.

Focusing on the comparison between $pN$- and $N$-FAM with standard QAOA, we recall that the number of free parameters in each ansatz is $pN + 2p$, $N + 2p$, and $2p$ respectively. Thus, $p \geq 4$ QAOA has more free parameters than $pN$- and $N$-FAM with $p=1$, and $p=5$ QAOA has more than $p = 2$ $N$-FAM. As can be seen, even when given more free parameters, standard QAOA does not achieve the level of performance seen for $pN$- and $N$-FAM. This demonstrates the benefit of not just more free parameters in an anstaz, but the informed choice of those free parameters to make more of Hilbert space accessible. FAM-QAOA does this by expanding the reach of the local mixer unitary.

\begin{figure}[t!]
    \centering
    \includegraphics[width=\columnwidth]{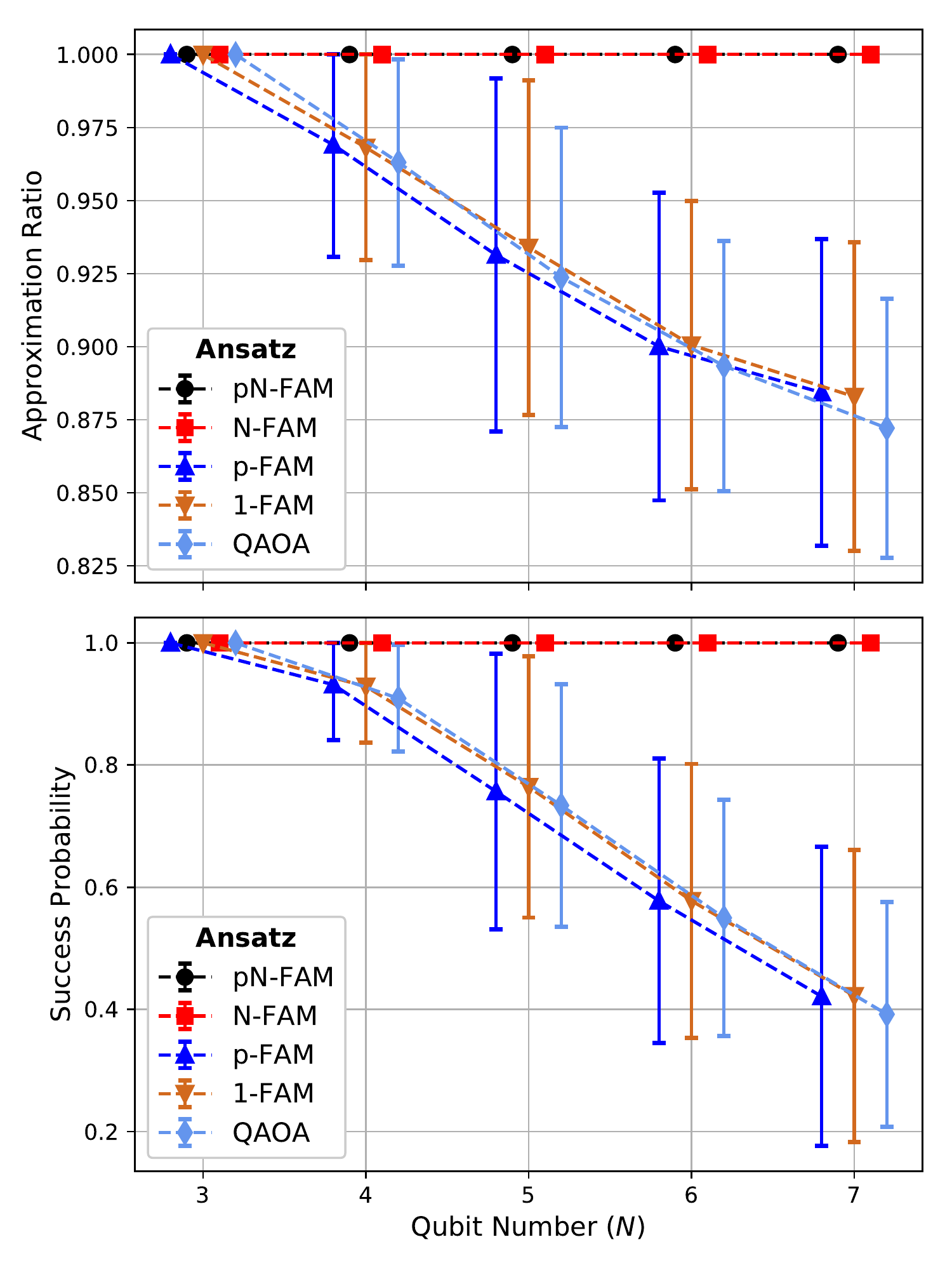}
    \caption{Average and error bars showing one standard deviation of the (upper panel) approximation ratio and (lower panel) success probability at MAXCUT as a function of the number of qubits ($N$) for all connected graphs for $N < 7$ and 100 connected graphs for $N=7$, with $p=2$, zero error, and 10 cold-starts.}
    \label{fig:fig3}
\end{figure}

Finally, for $p = 2$ and zero error we explore the performance as a function of qubit number, as shown in Fig.~\ref{fig:fig3} for all connected graphs on $N=3$ to $N=6$ qubits, and 100 connected graphs selected uniformly randomly for $N=7$. As we have consistently observed, $pN$- and $N$-FAM outperform all other ansätze, with no visible loss of AR or SP even at $N=7$. It is important to note that for all results presented in this section we have taken the best AR or SP from a number of repetitions with distinct initial conditions. If we consider the average performance over optimization repetitions, we do observe a loss of performance as $N$ increases for $pN$- and $N$-FAM, as shown in Appendix \ref{app:moresim}, though they still outperform all other ansätze.

\subsection{MAXCUT on a Neutral Atom Processor}

Existing quantum computing architectures will not generally be able to directly implement the problem and mixer unitaries necessary to execute QAOA. Instead, these unitaries must be decomposed into the native gate set of the architecture. An architecture with a universal gate set may implement any unitary as a sequence of native gate operations. Furthermore, two qubit operations cannot in general be applied to arbitrary pairs of qubits. Two qubit operations that cannot be performed due to the connectivity constraints of the architecture must be routed via the addition of SWAP gates.

In the case of the neutral atom processor at UW-Madison \cite{Graham2019}, there are three native gate operations. A phase gate implements a rotation about the Z-axis for a single qubit which is achieved by applying a Stark-shifting laser field to the qubit. A global microwave gate implements a rotation about an arbitrary axis in the X-Y plane on each qubit. This gate is global because the spacing between qubits is far less than the wavelength of the microwave field, and thus each qubit is rotated in the same way. The axis of rotation may be chosen by modifying the phase of the microwave field. Lastly, the two qubit controlled-Z (CZ) gate is implemented by the Rydberg blockade effect. Put together, these operations form a universal gate set which can be used to implement QAOA for the MAXCUT problem. Physical qubits are optically trapped in a square-grid lattice. When routing two qubit operations, nearest-neighbor connectivity is assumed when inserting SWAP operations.

We consider a noise model where each type of gate operation has distinct noise characteristics. After a global microwave gate is applied, each qubit experiences $T_1, T_2$ decoherence that drives the system towards the maximally mixed state. The decoherence is characterized by a $T_1$ time of 0.5 s and a $T_2$ time of 10 ms and each microwave gate is assumed to have a 25$\mu$s duration. Additionally, each microwave gate is characterized by a  single-qubit depolarizing channel with depolarizing probability $5\times 10^{-4}$ which is applied to each qubit after the $T_1, T_2$ decoherence channel. Only microwave gates are followed by $T_1, T_2$ decoherence because both Z-rotation and CZ gates have a significantly shorter duration than microwave gates (<1$\mu$s). After a Z-rotation gate, the target qubit experiences a single-qubit depolarizing channel with probability $10^{-3}$. Ideal CZ gates are replaced by the non-ideal unitary
\begin{equation}\label{eq:imperfect_cz}
    \begin{pmatrix}
    1 & 0 & 0 & 0 \\
    0 & \sqrt{1-E_1} & \sqrt{E_1}e^{i \phi} & 0\\
    0 & -\sqrt{E_1}e^{-i \phi} & \sqrt{1-E_1} & 0\\
    0 & 0 & 0 & -e^{i \delta}
    \end{pmatrix},
\end{equation}
where $E_1$ and $\phi$ parameterize an undesired rotation in the $\{\ket{01},\ket{10}\}$ subspace and $\delta$ characterizes the deviation from the ideal controlled phase accumulation of -1 \cite{Geller2013}. To first order, the state averaged gate fidelity $F$ can be related to the error parameters by \cite{Geller2013}
\begin{equation}\label{eq:first_order_fidelity}
1-F = \frac{2}{5} E_1 + \frac{3}{20} \delta^2.
\end{equation}
We assume an equal contribution from each of the terms in  Eq.~\eqref{eq:first_order_fidelity} and arbitrarily set $\phi=\pi/4$. In the simulations, the single qubit noise is kept constant while the CZ fidelity is allowed to vary except for the case of $F=1$, for which no noise is applied.

\begin{figure}[t!]
    \centering
    \includegraphics[width=\columnwidth]{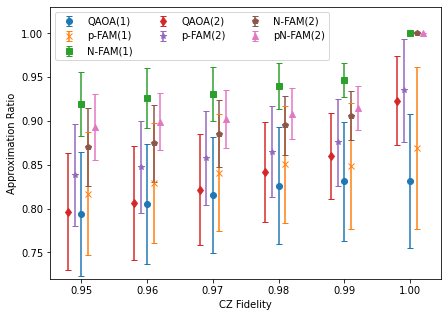}
    \caption{Approximation ratios achieved by each ansatz when taking the best of 10 repetitions versus CZ fidelity. Results are averaged over all 5-qubit connected graphs and error bars show one standard deviation of the performance across graphs. Markers evaluated at the same value of the CZ fidelity for different ansatz are offset slightly horizontally for ease of comparison.}
    \label{fig:bestofatlasfig}
\end{figure}

Fig.~\ref{fig:bestofatlasfig} shows the performance of natively compiled FAM-QAOA and standard QAOA circuits with respect to this noise model for various CZ fidelity values. Ten random initial conditions were generated for each ansatz and the same initial conditions were used for each fidelity. For each symbol, the best result from the ten repetitions was taken for each graph and averaged over all connected 5-qubit graphs. The number of functions calls was capped at 1500 per repetition for each ansatz.

As in the hardware agnostic case, the $N$- and $pN$- ansatz clearly outperform the rest. The FAM-QAOA versions with less free parameters still perform favorably compared to standard QAOA. For sufficiently high levels of noise, $p$-FAM at $p=1$ outperforms standard QAOA at $p=2$ despite having fewer variational parameters.

\section{Exact Solutions for Single Round FAM-QAOA}
\label{sec:exact}

For single round QAOA, one can constrain the class of problem Hamiltonians that can be solved exactly by examining both sides of the equation \cite{McClean2020}
\begin{align}
    e^{-i\gamma\hat{H}_Z}\ket{+}^{\otimes N} = \hat{U}^\dagger_M\ket{z^*}, \label{eqn:exactqaoa}
\end{align}
where $\ket{z^*}$ is the unique \footnote{This analysis works under the assumption that the ground-state is unique.} ground-state of $\hat{H}_Z$. Previous work has shown that for standard QAOA the family of problem Hamiltonians that can be solved in a single round have the form
\begin{align}
    \hat{H}_Z = \alpha\sum_{n=1}^{N}\hat{Z}_n + \frac{2\pi m}{\gamma}\left({\rm Many~body~Z~terms}\right),
\end{align}
for $m\in\mathbb{Z}$. The intuition behind this solution is that $\beta = \pi/4$ is fixed to ensure an equal probability superposition on the right-hand side, and then $\gamma$ is chosen such that $\alpha\gamma = \pi/4$. Any terms in $\hat{H}_Z$, even those that act nontrivially on multiple qubits, that have a pre-factor that is an integer multiple of $2\pi/\gamma$ will impart a trivial phase on the left-hand side of Eq.~\eqref{eqn:exactqaoa}.

For a similar analysis of FAM-QAOA, we rewrite Eq.~\eqref{eqn:exactqaoa} as
\begin{align}
    e^{-i\gamma\hat{H}_Z}\ket{+}^{\otimes N} = e^{-i\boldsymbol{\theta}.\mathbf{Z}}e^{i\beta\hat{H}_X}e^{i\boldsymbol{\theta}.\mathbf{Z}}\ket{z^*},
\end{align}
from which it is clear that $\beta = \pi/4$ will again be necessary. With this in mind, we obtain
\begin{align}
    e^{-i\left(\gamma\hat{H}_Z - \boldsymbol{\theta}.\mathbf{Z}\right)}\ket{+}^{\otimes N} = e^{i\theta_\Sigma}\exp\left(i\frac{\pi}{4}\sum_n\hat{Z}_n\right)\ket{+}^{\otimes N},
\end{align}
where
\begin{align}
   \theta_\Sigma =  \sum_n(-1)^{z^*_n}\theta_n,
\end{align}
is a phase factor corresponding to the sum of the free angles with a $\pm$ sign depending on the state of each qubit in the solution $\ket{z^*}$.

From this it is clear that we require
\begin{align}
    \exp\left(-i\left[\gamma\hat{H}_Z - \boldsymbol{\tilde{\theta}}.\mathbf{Z} - \theta_\Sigma\iden\right]\right) = \iden,
\end{align}
where $\boldsymbol{\tilde{\theta}} = \boldsymbol{\theta} + \pi/4$. Intuitively, as the above expression shows, the ability to choose $\theta_n$ for each qubit gives additional freedom for $\hat{H}_Z$, such that any Hamiltonian of the form
\begin{align}
    \hat{H}_Z = \sum_{n=1}^{N} \alpha_n\hat{Z}_n + \frac{2\pi m}{\gamma}\left({\rm Many~body~Z~terms}\right),
\end{align}
admits a solution with single round FAM-QAOA.

\section{Conclusion}
\label{sec:conc}

The results of this manuscript demonstrate that FAM-QAOA is a high-performing alternative to standard QAOA at small $p$ and $N$. We believe the improvement in performance is due to an increased ability to explore Hilbert space, which comes not just from having more free parameters, but from a designed choice of where additional freedom is given to the ansatz. This would indicate that such improvement in performance should extend into the large $p$ and $N$ regime, though we are unable to test this in numerical simulation. We note that FAM-QAOA is particularly good at achieving high success probability, which bodes well for applications where getting the exact ground state is more important than a high quality solution (as measured by the approximation ratio).

FAM-QAOA comes with a faster growth in free parameters than standard QAOA, and in particular it introduces a linear scaling with qubit number. However, we emphasize that FAM-QAOA was designed such that it is no less hardware-efficient than standard QAOA. With more free parameters one would expect an increase in the classical co-processing for optimization, and in the number of queries to the quantum processor. We find the latter to not be the case for small $p$ and $N$, and finding the right balance between classical computational cost and performance of a many-parameter ansatz is a problem for all variational quantum algorithms at large scale.

A key message of our results is that carefully choosing the additional freedom in an ansatz can be far more effective than adding more free parameters at random, or simply increasing circuit depth. To the latter point, we show that FAM-QAOA with smaller $p$ outperforms standard QAOA at larger $p$, which has great practical significance in the NISQ-era, where circuit depth (a function of $p$) is a tightly constrained resource. Exploring the full extent of the gains that can be achieved by FAM-QAOA on near-term hardware would be an interesting topic for future work, as would the study of what properties of modified mixers improve performance in QAOA-like ansatz.

\acknowledgements
The authors thank J.~Wurtz for providing the graph atlas of connected graphs used in this work. This work was supported by DARPA-ONISQ Contract No. HR001120C0068 and DOE Award DE-SC0019465.

\bibliography{QAOA}

\appendix

\clearpage

\section{Numerical Simulation Details}
\label{app:simdetails}

\subsection{Hardware Agnostic Simulations}

\begin{figure}[t!]
    \centering
    \includegraphics[width=\columnwidth]{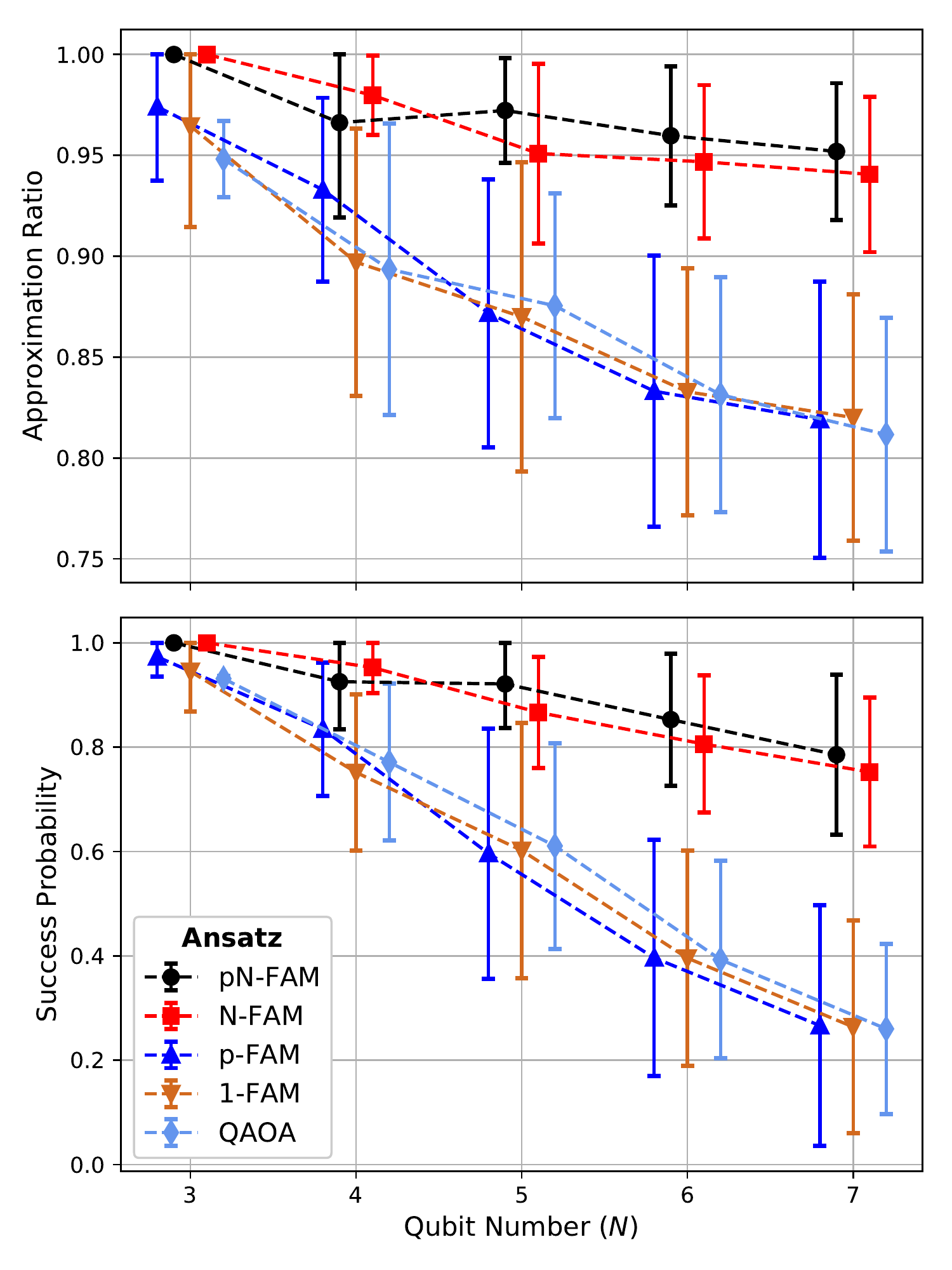}
    \caption{Average and error bars showing one standard deviation of the (upper panel) approximation ratio and (lower panel) success probability at MAXCUT as a function of the number of qubits ($N$) for all connected graphs for $N < 7$ and 100 connected graphs for $N=7$, with $p=2$ and zero error. This figure shows the average case performance over 10 repetitions with different initial conditions, while Fig.~\ref{fig:fig3} shows the best case performance.}
    \label{fig:figS2}
\end{figure}

The hardware agnostic MAXCUT numerical simulations were performed using custom made code for the quantum evolution written in {\tt Julia} \cite{Bezanson:2017aa}, which used {\tt Optim.jl} \cite{Mogensen2018} for the optimization. For the results in Figs.~\ref{fig:fig1} and \ref{fig:fig2}, the {\tt Particle Swarm} optimizer from {\tt Optim.jl}, based on the algorithm of Ref.~\cite{Zhan2009}, was used as it gave the most consistent results for all error models. For the results of Fig.~\ref{fig:fig3}, the {\tt BFGS} optimizer was used as it reduced simulation time and remained accurate for the zero error model. In all cases, the $N$- and $1$-FAM simulations used the scaled version of the ansatz, Eq.~\eqref{eqn:FAMQAOAscaled}.

For the results presented in the main text the number of function calls, i.e.~evaluations of the FAM-QAOA cost function by forward simulation of Eq.~\eqref{eqn:fullFAM}, was capped at $5001$, with a restriction of $1000$ iterations for the {\tt Particle Swarm} optimizer. For the {\tt BFGS} optimizer, the number of iterations and function calls was typically at least an order of magnitude less. For the fixed error and $\gamma$-dependent error models, $\phi = 0.1\pi$. For the qubit-dependent and qubit/$\gamma$-dependent error models (Fig.~\ref{fig:fig1}), $3$ error vectors $\boldsymbol{\phi}$, each of length $N=5$, were chosen with elements $\phi_n$ drawn uniformly randomly from $\left[0, 0.2\pi\right]$. The results presented are the average performance over these three error instances. For more information on performance with varying function calls, and as a function of error for the fixed error model see Appendix \ref{app:moresim}.

The main text results are the best case performance over a series of $3$ (Figs.~\ref{fig:fig1} and \ref{fig:fig2}) or $10$ (Fig.~\ref{fig:fig3}) repetitions of the FAM-QAOA optimization with different initial conditions. Fig.~\ref{fig:figS2} shows the AR and SP as a function of qubit number $N$, similar to Fig.~\ref{fig:fig3}, but with the crucial difference that Fig.~\ref{fig:figS2} shows average performance over 10 repetitions (i.e.~new optimizations with different initial conditions) instead of best case. As can be seen, $pN$- and $N$-FAM continue to outperform the other FAM-QAOA versions and standard QAOA, but compared to Fig.~\ref{fig:fig3} their average case performance degrades as a function of $N$. Thus, while best case performance was not impacted for $3\leq N \leq 7$, it appears the probability of finding a high quality solution, given a random initial condition, reduces as qubit number increases.

\subsection{Neutral Atom System Simulations}

\begin{figure}[t!]
    \centering
    \includegraphics[width=\columnwidth]{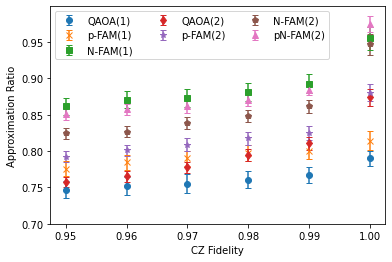}
    \caption{Approximation ratios achieved by each ansatz averaged over 10 repetitions versus CZ fidelity. Results are averaged over all 5-qubit connected graphs and error bars are 3$\sigma$ standard errors of the mean.}
    \label{fig:meanatlasfig}
\end{figure}

\begin{figure}[t!]
    \centering
    \includegraphics[width=\columnwidth]{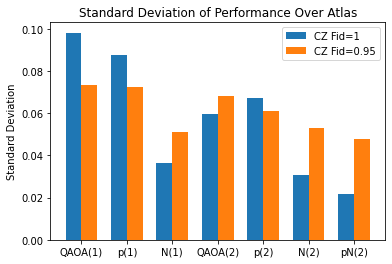}
    \caption{Standard deviation of the mean performance obtained for each graph for CZ Fidelity of 0.95 and 1.0.}
    \label{fig:atlas_variance}
\end{figure}

The simulations of quantum circuits matching the capabilities of neutral atom hardware were performed using the quantum simulation package {\tt Cirq} \cite{cirq_developers_2021_4586899}. Classical optimization of the variational parameters was performed using the {\tt Nelder Mead} optimizer in {\tt Scipy} \cite{Scipy}. The results in the main text were the best case performance over 10 repetitions of the FAM-QAOA optimization with different initial conditions. Fig.~\ref{fig:meanatlasfig} shows the average case performance for the same 10 repetitions.  The initial conditions were generated at random for each ansatz and were the same for each value of the CZ fidelity. The single qubit noise was kept the same for each value of the CZ fidelity except for 1, for which no noise was applied. The number of function evaluation calls per repetition was capped at 1500.

To estimate the uncertainty in the average case, we first calculated a standard error of the mean for each of the 21 5-qubit graphs. The result shown in the plot is the average of the means for the 21 graphs, and these 21 standard errors are propagated into a standard error for the mean of the average case performance over all graphs. The relative performance of each ansatz in the average case was similar to the best case behavior presented in the main text. It is also useful to consider the variation in performance over graphs. To calculate this, we take the standard deviation of the means for each graph as we did for the best case plots presented in the main text. The resulting standard deviation of performance over the atlas for CZ fidelities of 0.95 and 1.0 are shown in Fig.~\ref{fig:atlas_variance}. We find that the FAM-ansatz tends to have reduced variance in performance with respect to choice of graph compared to standard QAOA when evaluated with the same noise parameters. Thus, the FAM-ansatz not only improves the average case performance, but reduces the variance in performance.

\section{Additional Numerical Simulations}
\label{app:moresim}

\begin{figure}[ht]
    \centering
    \includegraphics[width=\columnwidth]{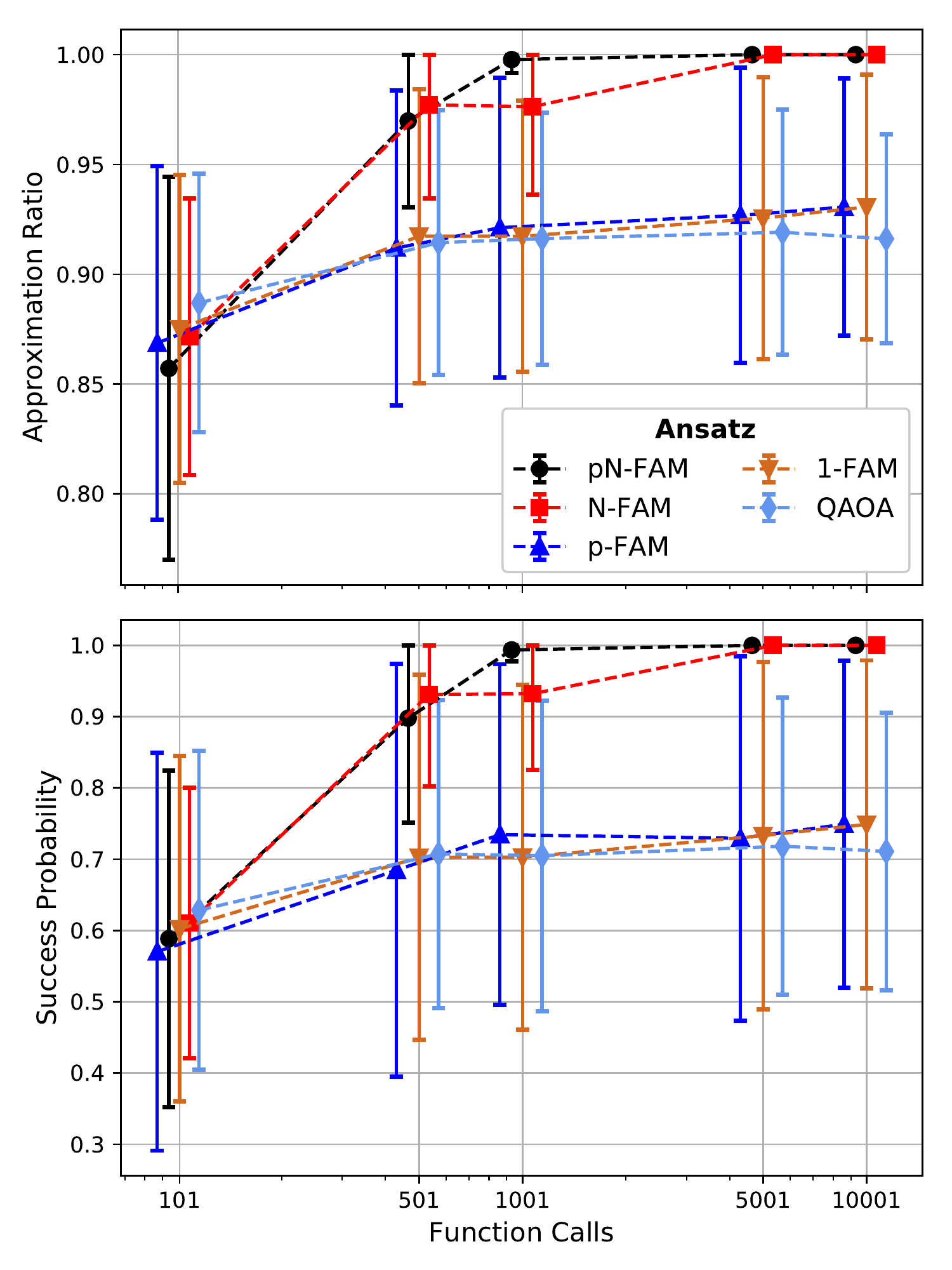}
    \caption{Average and error bars showing one standard deviation of the best-case (upper panel) approximation ratio and (lower panel) success probability at MAXCUT as a function of the number of function calls used by the optimizer, for all connected graphs with $N=5$, $p=2$, and zero error.}
    \label{fig:figS1}
\end{figure}

In this appendix we discuss supporting results to those shown in the main text. Fig.~\ref{fig:figS1} shows AR and SP for the FAM-QAOA versions and QAOA as a function of the number of function calls given to the {\tt Particle Swarm} optimizer. As can be seen, performance saturates around $5001$ function calls, which is the value used for the main text results. Note that these are best case results taken over three repetitions.

\begin{figure}[t!]
    \centering
    \includegraphics[width=\columnwidth]{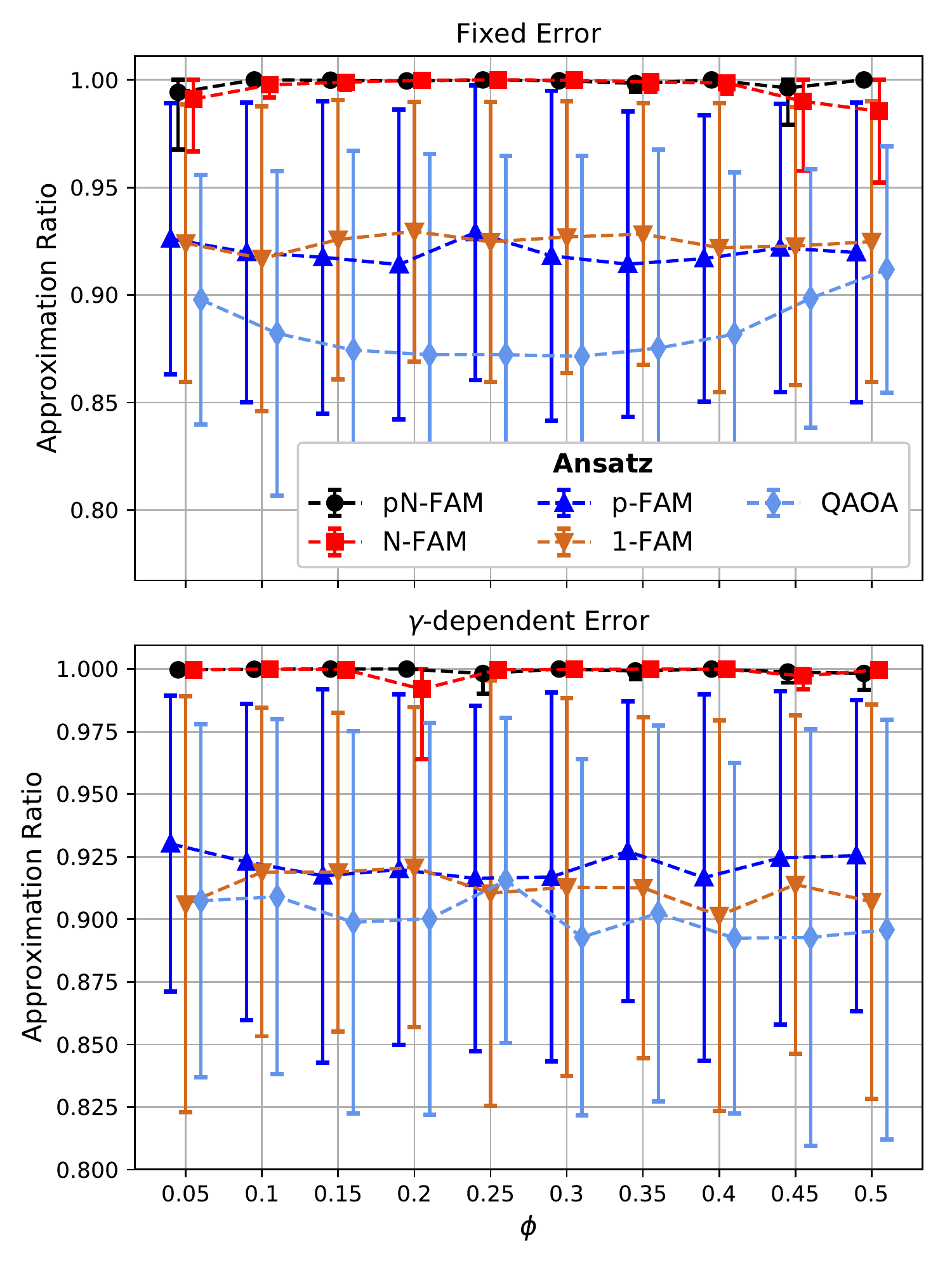}
    \caption{Average and error bars showing one standard deviation of the best-case approximation ratio for MAXCUT as a function of the Z-phase error magnitude $\phi$ for the (upper panel) fixed and (lower panel) $\gamma$-dependent error models, for all connected graphs with $N=5$ and $p=2$.}
    \label{fig:figS3}
\end{figure}

Fig.~\ref{fig:figS3} shows the AR as a function of the Z-phase error magnitude $\phi$ for the fixed error model, where $\phi_n^k = \phi$, and the $\gamma$-dependent error model where $\phi^k_n = \gamma_k\phi$. Performance for all the FAM-QAOA versions is roughly independent of $\phi$ for both error models, while QAOA shows weak $\phi$-dependence for the fixed error model.

\end{document}